\documentclass[english]{article}
\usepackage[T1]{fontenc}
\usepackage[latin9]{inputenc}
\usepackage{geometry}
\usepackage{fancyhdr}
\usepackage{refstyle}
\usepackage{graphicx}

\makeatletter

\AtBeginDocument{\providecommand\figref[1]{\ref{fig:#1}}}
\AtBeginDocument{\providecommand\secref[1]{\ref{sec:#1}}}
\AtBeginDocument{\providecommand\eqref[1]{\ref{eq:#1}}}
\AtBeginDocument{\providecommand\subref[1]{\ref{sub:#1}}}
\RS@ifundefined{subref}
  {\def\RSsubtxt{section~}\newref{sub}{name = \RSsubtxt}}
  {}
\RS@ifundefined{thmref}
  {\def\RSthmtxt{theorem~}\newref{thm}{name = \RSthmtxt}}
  {}
\RS@ifundefined{lemref}
  {\def\RSlemtxt{lemma~}\newref{lem}{name = \RSlemtxt}}
  {}

\usepackage{lastpage} 
\makeatletter
\let\ps@plain\ps@fancy 
\makeatother
\usepackage{hyperref}
\usepackage{cite} 

\makeatother

\usepackage{babel}
\begin{document}

\title{\textbf{Smelling out Code Clones:\\Clone Detection Tool Evaluation
and\\Corresponding Challenges}}

\author{Rachel Gauci\\
The University of Edinburgh\\
\href{mailto:s1402642@sms.ed.ac.uk}{s1402642@sms.ed.ac.uk}}
\maketitle
\begin{abstract}
Software clones have been an active area of research for the past
two decades. However, although numerous clone detection tools are
now available, only a small fraction of the literature has focused
on tool evaluation, and this is in fact still an open problem. This
is mostly due to the fact that standard information retrieval metrics
such as recall and precision require a priori knowledge of clones
already in the system. Detection tools also typically have a large
number of parameters which are difficult to fine-tune for optimal
performance on a particular software system, and different outputs
produced by different tools add to the complexity of comparing one
tool to another. In this review, we further explore the reasons why
tool evaluation is still an open challenge, and present the current
tools and frameworks targeted at mitigating these problems, focusing
on the current standard benchmarks used to evaluate modern clone detection
tools, and also presenting a recent method aimed at finding optimal
tool configurations.

\textit{The research work disclosed in this publication is funded
by the MASTER it! Scholarship Scheme (Malta). The scholarship is part-financed
by the European Union - European Social Fund (ESF) under Operational
Programme II - Cohesion Policy 2007-2013, ``Empowering People for
More Jobs and a Better Quality of Life''.}
\end{abstract}

\section{Introduction and Motivation\label{sec:Introduction-and-Motivation}}

Software clones are similar code fragments found in a single codebase,
and are typically the result of developers' ``cut-copy-paste-adapt
techniques'' \cite{rattan2013software}. Clones have been identified
as giving off a \textit{bad smell} \cite{beck1999bad} in code, and
are generally considered to be a bad programming practice. Code clones
lead to bug propagation, since a bug in a single code fragment will
be replicated in all clones of that fragment, causing an increase
in both difficulties and expenses related to project maintenance.
Code clones also unnecessarily bloat the size of a software system,
causing a strain on resources, and it is therefore desirable to eliminate
or at least limit the number of clones in a system.

\begin{figure}[t]
\begin{centering}
\includegraphics[scale=0.5]{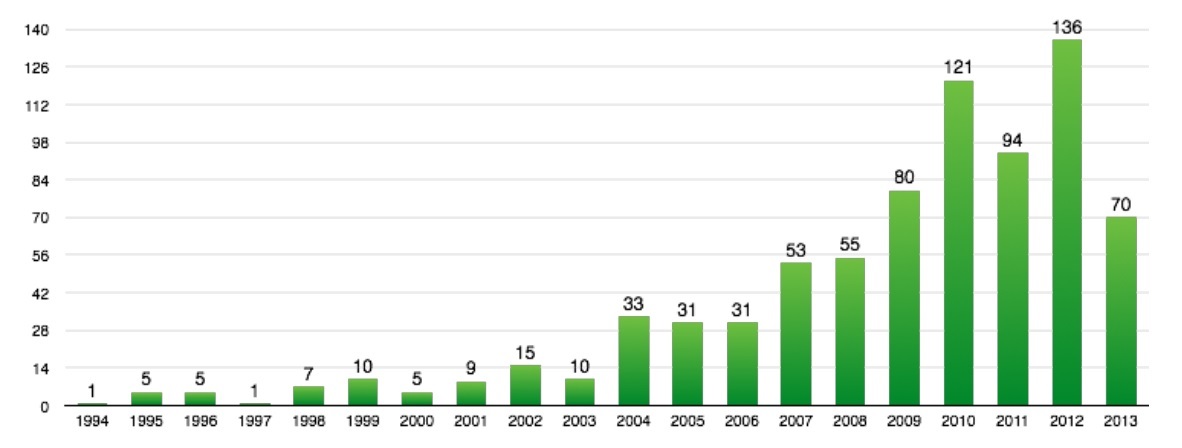}
\par\end{centering}

\protect\caption{Yearly number of distinct authors contributing to clone research \cite{roy2014vision}.
\label{fig:Yearly-number-of} }
\end{figure}

\begin{figure}[t]
\begin{centering}
\includegraphics[scale=0.5]{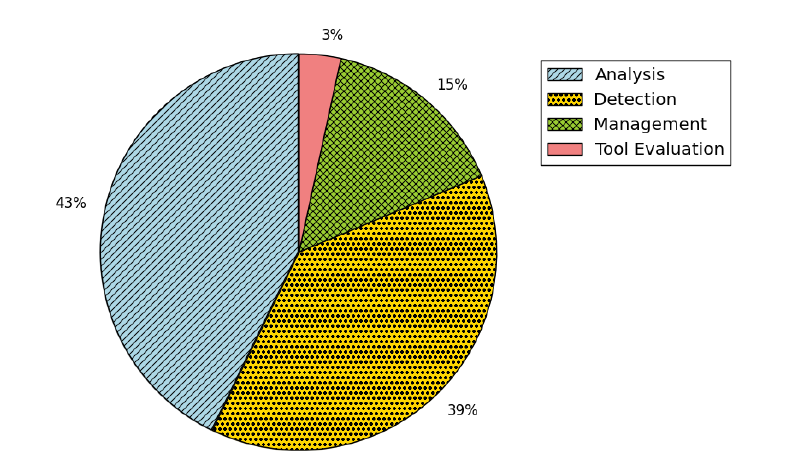}
\par\end{centering}

\protect\caption{Proportion of publications in each category over the period 1994-2013
\cite{roy2014vision}.\label{fig:Proportion-of-publications}}

\end{figure}

Software clones emerged as a research area $20$ years ago in $1994$
\cite{roy2014vision} and, as can be seen from \figref{Yearly-number-of},
the field has lately experienced a significant increase in interest,
with more and more authors contributing to the literature in recent
years. From \figref{Proportion-of-publications}, we see that research
ranges across four sub-areas \cite{roy2014vision} with varying degrees
of popularity: clone analysis, clone detection, clone management and
tool evaluation. In particular we note that most of the studies have
focused on clone detection and clone analysis, two sub-areas which
are concerned with the development of tools and techniques for clone
identification, and the analysis of clone traits and features, such
as reasons for their existence and their effects on codebases and
project maintenance. Research in clone detection has resulted in more
than $70$ \cite{svajlenko2014evaluating} detection tools being currently
in existence, however only $3\%$ \cite{roy2014vision} of the literature
has been dedicated to the evaluation of these tools.

Tool evaluation is important for two main reasons. Firstly, when a
new detection tool is made available, we would like to be able to
evaluate its performance, both by itself and in relation to other
detection tools; perhaps the new tool is better at identifying a particular
type of clones, or has a lower false positive rate, or less demanding
space and time requirements. Secondly, tool evaluation plays an important
role in  tool selection: given a particular software system, written
in a particular programming language, having a particular size, and
suspected to contain certain types of clones, we would like to know
which detection tool would be the most suitable for the clone detection
task, without having to try the tools out one by one and compare or
verify their output.

Tool evaluation has however proved to be a very difficult task. To
this day, evaluation still remains an open challenge \cite{roy2014vision},
and in this review we explore the challenges in this sub-area of clone
research. Currently, only two benchmarks exist for tool evaluation
\cite{roy2014vision}: Bellon's Framework \cite{bellon2007comparison}
and Roy et al.'s Mutation \& Injection Framework \cite{roy2009mutation,svajlenko2013mutation},
and this review focuses on their limitations while also presenting
the recent EvaClone tool \cite{wang2013searching}, which aims to
mitigate the tool configuration problem threatening the validity of
tool comparison experiments.

\section{Clone Detection Tool Evaluation}

\subsection{Challenges in Tool Evaluation\label{sub:Challenges-in-Tool}}

Given a software system and a resultant set of clones detected by
a particular tool, we would ideally be able to come up with quantitative
measures for the tool's coverage, whether it was able to find all
the clones in the system, and the tool's accuracy, whether any of
the reported clones were actually false positives. As mentioned in
\secref{Introduction-and-Motivation}, this kind of tool evaluation
is not a trivial task, and this is due to three main reasons, which
we outline below.

\subsubsection{Lack of Reference Corpora}

The standard metrics in the field of Information Retrieval, \textbf{recall}
and \textbf{precision}, require a priori knowledge of the clones already
in the system. Recall is concerned with a tool's coverage. For a detection
tool $T$ and a software system $S$, recall can be calculated as
in \eqref{recall}, where a recall value of $1$ would indicate that
the tool was able to detect all clones in the system.
\begin{equation}
\mathrm{recall}_{T,S}=\frac{\mathrm{number\, of\, detected\, clones}}{\mathrm{total\, number\, of\, clones\, in\, system}}\label{eq:recall}
\end{equation}
Precision, on the other hand, is concerned with accuracy, and for
a detection tool $T$ and software system $S$, can be calculated
as in \eqref{precision}, where a value of $1$ would indicate that
all reported clones where indeed true positives
\begin{equation}
\mathrm{precision}_{T,S}=\frac{\mathrm{number\, of\, actual\, clones\, from\, reported\, clones}}{\mathrm{total\, number\, of\, reported\, clones}}\label{eq:precision}
\end{equation}
The two metrics work at odds with each other, since relaxing a tool's
similarity measure will increase its recall while sacrificing its
precision, and vice-versa. It is easy, for example, to come up with
a tool having $100\%$ recall by simply reporting all possible combinations
of code fragments as clones, but such a system would have a very low
precision value and would, in effect, be useless. Therefore a good
clone detection tool should have both high recall and high precision.
The main problem with these metrics is, however, that they cannot
be calculated without first identifying the clones in the system.

As pointed out by Bellon et al. \cite{bellon2007comparison}, there
are three obvious, naïve ways in which a reference corpus can be constructed
from the candidate clones reported by a number of detections tools:
\begin{enumerate}
\item \textbf{\label{enu:union-of-candidates:}union of candidates}: using
this as a reference corpus, we would end up with a large number of
false positives, as it is enough for one tool to classify two clone
fragments as clones for the pair to be included in the reference corpus.
Not only this, but this approach would result in a precision of $1$
for all tools, which is not in any way representative of the tools'
true precisions, but merely a result of an inaccurate definition of
what constitutes a reference clone.
\item \textbf{\label{enu:intersection-of-candidates}intersection of candidates
found by all tools}: using the intersection instead of the union would
have the opposite effect, leading to an inaccurate recall value of
$1$ for all tools, and a large number of true negatives, since failure
of a single detection tool to identify an actual clone would lead
to that clone being excluded from the reference corpus.
\item \textbf{intersection of candidates found by at least $N$ Tools}:
this approach is essentially a compromise between \ref{enu:union-of-candidates:}
and \ref{enu:intersection-of-candidates} above, but is still not
a good solution, as no matter what value is chosen for $N$, $N$
tools might still report a false positive, while $N-1$ tools identify
a true positive \cite{bellon2007comparison}.
\end{enumerate}
Bellon's Framework \cite{bellon2007comparison} and the Mutation \&
Injection Framework \cite{roy2009mutation,svajlenko2013mutation}
address the problems of a missing reference corpus by constructing
a corpus of reference clones using two very different methodologies.
In Bellon et al.'s approach \cite{bellon2007comparison}, $2\%$ of
the $325,935$ clones identified by $6$ different clone detectors
on $8$ large C and Java software systems were manually ``oracled''
by Stefan Bellon, and the ones accepted by Bellon as actual clones
were included as clones in the reference corpus. In Roy et al.'s approach
\cite{roy2009mutation,svajlenko2013mutation}, mutation operators
are used to automatically generate synthetic clones from randomly-selected
code fragments in the system, where the mutation operators aim to
imitate the copy, paste and modify operations performed by developers
which lead to real code clones in software systems. The artificial
clones generated by the mutation operators are then injected into
the software systems and used as the reference corpus for evaluation.
We discuss the advantages and limitations of the approaches adapted
by these two benchmarks in \subref{Evaluation-Benchmarks}.

\subsubsection{The Tool Configuration Problem}

Each detection tool has several different parameters and configuration
options and, for optimal performance of the tool, these need to be
fine-tuned according to the software system being analyzed. This problem
is well recognized in clone literature, with as many as $113$ papers
commenting on the effects that tool configurations can have on empirical
evaluations, $57$ of which have a dedicated section discussing \textit{threats
to validity} \cite{wang2013searching}. This ``confounding configuration
choice problem'', as coined by Wang et al. \cite{wang2013searching},
threatens the validity of evaluation results, since differences observed
in tool performance might not be related to properties of the clones
or the tools themselves, but may be the result of the configurations
and parameters used in that particular experiment.

EvaClone%
\footnote{We use the term ``EvaClone'' to represent both the \textit{EvaClone}
and \textit{CloudEvaClone} tools, were the latter is simply a cloud-based,
parallelized version of the original \textit{EvaClone} desktop application
\cite{wang2013searching}.%
} \cite{wang2013searching} aims to mitigate this confounding configuration
problem. Given a set of detection tools and software systems, it makes
use of a genetic algorithm to search the space of all possible tool
configurations, in order to find the optimal configurations for each
tool. The default fitness function aims to maximize tool agreement,
i.e. maximize the size of the intersection of clone candidates, but
the EvaClone framework allows for any choice of fitness function since
the function itself is a parameter, and we further review this framework
in \subref{The-Tool-Configuration}.

\subsubsection{Incompatible Tool Output Formats}

Different clone detectors output their results in different formats,
such as HTML, XML and plain-text, and this complicates the process
of head-to-head empirical evaluations. Not only this, but the clone
results differ in terms of clone size and granularity, and the clones
types identified by the tool and their degree of similarity. We do
not focus on this problem in this review, since the evaluations carried
out for Bellon's Framework \cite{bellon2007comparison}, the Mutation
\& Injection Framework \cite{roy2009mutation,svajlenko2013mutation}
and EvaClone \cite{wang2013searching} have managed to avert this
problem by transforming and normalizing the outputs of the evaluated
tools, but we acknowledge that it does indeed present a challenge
to the evaluation of different tools.

Proposed formats aimed at standardizing tool output include Harder
and Göde's \textit{Rich Clone Format} (RCF) \cite{harder2011efficiently}
and Duala-Ekoko et al.'s Clone Region Descriptor (CRD) \cite{duala2010clone}.
Kapser et al. \cite{kapser2012common} also presented a draft proposal
for the setup of a unified clone model, and set up an online wiki
at \url{http://www.softwareclones.org/ucm} \cite{kamiya2012wiki}
for discussion and further proposals, although we noted that the discussion
forums on the wiki are unfortunately inactive.

\subsection{Limitations of Current Evaluation Benchmarks\label{sub:Evaluation-Benchmarks}}

Bellon's Framework \cite{bellon2007comparison} consists of a set
of reference clones constructed by manual verification of the clone
candidates reported by $6$ different tools which participated in
an evaluation experiment in 2002. These tools represented the state-of-the-art
at the time, including Dup \cite{baker1995finding,baker1997parameterized}
and CloneDR \cite{baxter1998clone,baxter2004dms}, and an early version
of CCFinder \cite{kamiya2002ccfinder}, the current, most widely-cited
detection tool in clone literature \cite{rattan2013software}.

The outcome of the experiment presented several interesting results
concerning the participating tools, including indications that token-based
and text-based tools have higher recall, while tree-based tools have
higher precision \cite{bellon2007comparison}.%
\footnote{A good explanation of these and other clone detection techniques can
be found in Rattan et al.'s literature survey \cite{rattan2013software},
together with examples of tools using each technique and an overview
of empirical comparison studies.%
} We note however that the experimental setup suffers from a number
of limitations, and believe these should be taken into account when
using the framework for evaluation. Our first concern is that the
reference corpus was built by a single person, in this case Stefan
Bellon, and this could have introduced unknown bias into the corpus.
Different developers have different tolerance for clones, and as pointed
out by Svajlenko et al. \cite{svajlenko2013mutation} and Wang et
al. \cite{wang2013searching}, the reliability of human judges is
questionable, as even expert judges often disagree on whether two
code fragments are cloned or not. Even in the case where human intervention
is necessary, we suggest that bias can perhaps be eliminated or at
least reduced by including more than one individual in the corpus-building
process.

Another limitation of Bellon's framework is that only $2\%$ of the
clone candidates were ``oracled'' because of time limitations, as
the $325,935$ candidates took Bellon $77$ hours to classify \cite{bellon2007comparison}.
As a countermeasure, two evaluation experiments were conducted, one
after each $1\%$ of the candidates were classified, and the authors
show that the distribution of the detected clones was similar after
both experiments, where the number of candidate clones per tool approximately
doubled after the second experiment.

Bellon et al.'s experiment \cite{bellon2007comparison} also involved
the injection of $50$ type 1, 2 and 3 clones into the software systems,
but the detection tools were only able to locate $43$ of them. This
brings into question the reliability of using the tools' results themselves
as a reference corpus, since actual clones might go undetected and
will therefore not be taken into account during the evaluation process.
The corpus could also be biased towards the detection tools used to
create it, putting other tools at a disadvantage, and the results
of a recent study by Svajlenko et al. \cite{svajlenko2014evaluating}
do indeed seem to suggest that this is the case. This study \cite{svajlenko2014evaluating}
involved a comparison of the performance of two versions of CCFinder
\cite{kamiya2002ccfinder,kamiya2008ccfinderx}: the one used in the
original experiment in 2002 \cite{kamiya2002ccfinder} and the more
recent CCFinderX \cite{kamiya2008ccfinderx}. The results of the study
\cite{svajlenko2014evaluating} show that, according to Bellon's Framework
\cite{bellon2007comparison}, the recall for the newer version is
worse for almost all three types of clones on both Java and C systems.
This brings to light what is possibly the current biggest limitation
of Bellon's Framework \cite{bellon2007comparison}: its corpus was
built from a number of tools which were contemporary to 2002, and
it might not be sufficient for evaluating the performance of modern
tools \cite{svajlenko2014evaluating}.

Roy et al.'s Mutation \& Injection Framework \cite{roy2009mutation,svajlenko2013mutation}
avoids this problem by creating artificial clones from original subject
systems through the use of mutation operators, and recent work \cite{svajlenko2013mutation}
on the original framework \cite{roy2009mutation} has generalized
the framework for use with any clone detection tool. The automatic
generation of clones also avoids the problem of human bias in the
corpus generation process, and the framework was in fact the first
attempt at a fully automated evaluation process \cite{roy2009mutation}.
However, as pointed out by Svajlenko et al. \cite{svajlenko2014evaluating},
this is also the framework's major ``threat to validity'', since
the artificial clones generated through mutation analysis might not
be representative of actual clones generated by developers.

For a more thorough comparison between these two frameworks, we refer
the reader to Svajlenko et al.'s recent study \cite{svajlenko2014evaluating}
comparing the recall of $11$ modern clone detection tools using both
frameworks, and comparing them to expected recall values. The Mutation
\& Injection Framework \cite{roy2009mutation} performs better overall
\cite{svajlenko2014evaluating}, although we would have also liked
to see precision measures included in the study since these metrics
should be taken into account together, as discussed in \subref{Challenges-in-Tool}.
We also note that the authors acknowledge the fact that their expectations,
although well-researched, could contain inaccuracies, and we also
add that they could contain an element of bias, since the authors
of the study \cite{svajlenko2014evaluating} are also the creators
of the Mutation \& Injection Framework \cite{roy2009mutation,svajlenko2013mutation}.

\subsection{Addressing The Tool Configuration Problem \label{sub:The-Tool-Configuration}}

As outlined in \subref{Challenges-in-Tool}, the tool configuration
problem is well acknowledged but unfortunately not well addressed
\cite{wang2013searching}. During Bellon et al.'s experiment \cite{bellon2007comparison},
the evaluation was split into two parts: \textit{mandatory}, using
the tools' default configurations, and \textit{voluntary}, where the
tools' authors could optionally submit the tool with its configurations
fine-tuned for the subject systems. However, only two of the authors
submitted their tool for the voluntary run. In Svajlenko et al.'s
recent study evaluating modern clone detection tools \cite{svajlenko2014evaluating},
the default configurations were used, and this is in fact included
in the paper as another ``threat to validity''.

Wang et al. \cite{wang2013searching} show that this is a recurring
problem in clone literature, where a large proportion of the literature
simply uses the default configuration settings for evaluation. Using
the optimal tool configurations, as determined by EvaClone \cite{wang2013searching},
the fitness values for different clone detectors on Java and C systems
was improved by up to $21.9\%$ and $10.6\%$, respectively, showing
that tool configuration can have a significant effect on the tool's
performance, and is indeed something to be taken into consideration
when carrying out comparative studies.

However, similar to what we have observed in other examples of clone
literature, EvaClone also comes with \textit{threats to validity}
\cite{wang2013searching}. Firstly, the subject systems used in the
tool's evaluation \cite{wang2013searching} were all open source,
and the authors admit that the results might not be representative
of other types of software systems. Secondly the tool's fitness function
presents another configuration problem in itself. The default fitness
function aims to maximize the agreement between tools, and our main
concern is that such a metric can inflate the recall value, as explained
in \subref{Challenges-in-Tool}, since this is essentially another
spin on using the intersection of clone candidates as a reference
corpus \cite{bellon2007comparison}. In fact, when Wang et al. \cite{wang2013searching}
used EvaClone's optimal tool configurations to evaluate performance
on the \textit{psql} and \textit{swing} systems using Bellon's Framework
\cite{bellon2007comparison}, they found that their fitness function
favours recall over precision, supporting our hypothesis that a different
fitness function might be more appropriate. The work on EvaClone and
corresponding results \cite{wang2013searching} however do confirm
that empirical studies are potentially flawed in the way evaluation
has been performed, and it would be very interesting to see how the
results from Svajlenko et al.'s thorough study of modern tool evaluation
\cite{svajlenko2014evaluating} would be impacted through the use
of optimal configurations.

\section{Conclusion and Future Work}

Through this review, we have highlighted the main challenges involved
in the open problem of performing clone detection tool evaluation,
and presented the current, state-of-the-art frameworks aimed at mitigating
these challenges: Bellon's Framework \cite{bellon2007comparison}
and the Mutation \& Injection Framework \cite{roy2009mutation,svajlenko2013mutation},
which have been used to generate reference corpora to enable the calculation
of evaluation metrics, and the EvaClone framework, which can be used
for automatic determination of optimal tool configurations, based
on a pre-determined fitness function.

Despite these being the state-of-the-art, all frameworks have been
shown to contain several limitations, but software clones are a very
active research area \cite{roy2014vision}, and we are therefore hopeful
that further progress will be made in the sub-area of tool evaluation.
Svajlenko et al. \cite{svajlenko2014evaluating} have confirmed that
their priorities for future work involves updating Bellon's Framework
\cite{bellon2007comparison} with clones detected by modern tools,
while Roy et al. \cite{roy2014vision} suggest that significant progress
could be made if research teams actually start exchanging data for
reasons other than simply benchmarking their tools.

Clone detection is also being applied to areas other than software,
such as clones in UML domain models \cite{storrle2010towards} and
MATLAB/Simulink models \cite{pham2009complete}. Other research, such
as that by Rahman et al. \cite{rahman2012clones}, focuses on the
advantages of having clones in a software system, and suggests that
clones might not be a \textit{bad smell} after all \cite{rahman2012clones},
although we note that this study also comes with its own \textit{threats
to validity} section, something which is unfortunately still very
common in clone literature.

However, despite all these difficulties, clone management is gathering
momentum in the industry. Several detection tools are now available
as plug-ins for the Eclipse IDE, and clone management features have
also been added to Microsoft Visual Studio \cite{roy2014vision},
and so we are hopeful that, with increasing interest in both the research
and industrial communities, new approaches to the problem of tool
evaluation will be proposed, so that the current limitations can be
addressed, enabling comparative, evaluation experiments to be carried
out without any further threats to validity.

\bibliographystyle{plain}
\bibliography{references}

\end{document}